\documentclass[conference]{IEEEtran}
\IEEEoverridecommandlockouts      
\usepackage{cite}
\usepackage{threeparttable}
\usepackage{adjustbox}
\usepackage{amsmath,amssymb,amsfonts}
\usepackage{algpseudocode}
\usepackage{algorithm}
\usepackage{graphicx}
\usepackage{textcomp}
\usepackage{hyperref}
\usepackage{xcolor}

\ifCLASSINFOpdf
\else
\fi

\begin{document}
%





\title{Interference-Aware PMI selection for MIMO systems in an O-RAN scenario \\
\thanks{Telecom Italia sponsors this project.}
}
\author{\IEEEauthorblockN{1\textsuperscript{st} Rawlings Ntassah}
\IEEEauthorblockA{\textit{DISI} \\
\textit{University of Trento}\\
Trento, Italy \\
 \href{mailto:rawlings.ntassah@unitn.it}{rawlings.ntassah@unitn.it}}
\and
\IEEEauthorblockN{2\textsuperscript{nd} Gian Michele Dell'Aera}
\IEEEauthorblockA{\textit{Research and Innovation} \\
\textit{Telecom Italia}\\
Torino, Italy \\
\href{mailto:gianmichele.dellaera@telecomitalia.it}{gianmichele.dellaera@telecomitalia.it}}
\and
\IEEEauthorblockN{3\textsuperscript{rd} Fabrizio Granelli}
\IEEEauthorblockA{\textit{DISI} \\
\textit{University of Trento}\\
Trento, Italy \\
\href{mailto:fabrizio.granelli@unitn.it}{fabrizio.granelli@unitn.it}}

}


\maketitle

\begin{abstract}
The optimization of Precoding Matrix Indicators (PMIs) is crucial for enhancing the performance of 5G networks, particularly in dense deployments where inter-cell interference is a significant challenge. Some approaches have leveraged AI/ML techniques for beamforming and beam selection, however, these methods often overlook the multi-objective nature of PMI selection, which requires balancing spectral efficiency (SE) and interference reduction. This paper proposes an interference-aware PMI selection method using an Advantage Actor-Critic (A2C) reinforcement learning model, designed for deployment within an O-RAN framework as an xApp. The proposed model prioritizes user equipment (UE) based on a novel strategy and adjusts PMI values accordingly, with interference management and efficient resource utilization. Experimental results in an O-RAN environment demonstrate the approach's effectiveness in improving network performance metrics, including SE and interference mitigation.
\end{abstract}


%
\IEEEpeerreviewmaketitle

\section{Introduction}
\footnote{This paper has been accepted for publication in the 2025 IEEE 11th International Conference on Network Softwarization (NetSoft). IEEE will publish the final version of this work as: R. Ntassah; G. M. Dell'Aera; F. Granelli "Interference-Aware PMI selection for MIMO systems in an O-RAN scenario", Proc. 2025 IEEE 11th International Conference on Network Softwarization (NetSoft),  Budapest, Hungary, June 2025.} 

\IEEEPARstart{T}{he} advent of 5G networks has revolutionized wireless communication by enabling enhanced mobile broadband, ultra-reliable low-latency communication, and massive machine-type communications \cite{zhang2019towards}. However, the dense deployment of these networks introduces significant challenges, particularly in managing inter-cell interference, which can severely impact network performance and user experience.

Beamforming has been proposed as a key technology to mitigate connectivity issues between user equipment (UE) and the network by providing high-directional beams with reduced interference \cite{rao20215g}. Various beamforming techniques aim to optimize the Signal-to-Interference-plus-Noise Ratio (SINR) while minimizing interference and power consumption. One crucial aspect of optimizing 5G networks is the selection of Precoding Matrix Indicators (PMIs), which are essential for effective beamforming. Given the complexity of beam selection and management, sophisticated techniques are required to address these challenges.

In 5G systems, the base station (BS) transmits a channel state reference signal (CS-RS) to the UE, which then estimates the channel conditions based on a predefined codebook and feeds back the channel state information (CSI). The CSI feedback can include the channel quality indicator (CQI) and rank indicator (RI) for open-loop spatial multiplexing, or CQI, RI, and PMI for closed-loop configurations, as outlined in \cite{3gpp38213}.

Recent advancements in artificial intelligence (AI) and machine learning (ML) have shown promise in addressing the stochastic nature of network conditions. Studies such as \cite{8938771} and \cite{10224533} have explored various AI/ML techniques for beamforming, power control, and interference management. With the evolution of Radio Access Network (RAN) architectures, such as O-RAN, these models can be integrated into the RAN Intelligent Controller (RIC) to enhance optimization.

The O-RAN framework facilitates the deployment of intelligent models and AI/ML techniques through its RIC, which supports xApps and rApps for optimizing RAN functionality. This work introduces a reinforcement learning (RL)-based xApp for PMI selection and interference minimization. The proposed system utilizes a centralized controller to manage PMI selection within a network using predefined Type I codebooks. The central controller collects CSI from all BSs and sends control messages back to them, enabling coordinated optimization.

The primary contribution of this work is the development of an interference-aware PMI selection method for MIMO systems. We propose an Interference-Aware Advantage Actor-Critic (Inter-A2C) RL model that prioritizes UEs based on a novel strategy and assigns new PMI values accordingly. The reward function incorporates both inter-cell interference and resource utilization to optimize system performance. We implemented and tested this scheme as an xApp, demonstrating its effectiveness in improving performance metrics.

The organization of this paper is outlined as follows: Section II reviews related work, while Section III details our system model and the formulation of the problem. In Section IV, we analyze performance and analysis, and finally, Section V presents our conclusions. 

\section{Related work}
The optimization of PMIs in 5G networks has been extensively researched due to its significant impact on network performance. With the availability of large-scale data, AI/ML techniques have emerged as a promising solution for optimizing MIMO systems. For instance, Unlercsen et al. \cite{unlercsen2016direction} investigated the prediction of the angle of arrival (AoA) using received signals and their associated SINR. In contrast, the authors in \cite{9259040} proposed a two-stage beamforming technique that does not rely on AoA information, focusing instead on minimizing interference while improving SINR.

Beam selection is equally crucial for effective network performance. Wei-Cheng Kao et al. \cite{kao2018ai} developed an AI model to learn millimeter-wave (mmWave) channel characteristics and select beams based on predefined codebooks, using Rosenbrock search for optimal beam selection. Similarly, the authors in \cite{8114345} proposed a beam selection method utilizing out-of-band sub-6 GHz information for mmWave channel estimation. Min Soo S. et al. \cite{9034044} explored deep learning methods for beam selection using the sub-6 GHz power delay profile (PDP). Furthermore, \cite{9681382} presented a lightweight neural network (NN) model leveraging side information from LiDAR sensors for beam selection in both line-of-sight (LoS) and non-line-of-sight (NLoS) scenarios.

Maximizing SINR often involves designing effective beamforming codebooks. Early works such as \cite{love2003grassmannian} discussed the design of Grassmannian beamforming codebooks. For high-frequency applications, the studies in \cite{5379063} explored codebook design for mmWave systems, highlighting the potential of high-frequency ranges for efficient beamforming.
Interference management remains a critical challenge in dense 5G deployments. Zhang et al. \cite{zhang2018joint} proposed a coordinated beamforming approach to mitigate inter-cell interference in heterogeneous networks. Björnson et al. \cite{bjornson2017massive} introduced a pilot contamination precoding method to address interference in massive MIMO systems. A low-complexity PMI selection method using Type II codebook prediction was presented in \cite{9605010}.

With the introduction of the Open Radio Access Network (O-RAN) paradigm, new opportunities for optimising beamforming have emerged. The study in \cite{9762865} demonstrated the application of cell-free massive MIMO for energy efficiency and improved SE using O-RAN's Radio Units (RU) and Distributed Units (DU). Ryan M. et al. \cite{dreifuerst2023ml} employed a neural network to optimize codebooks in sub-6 GHz networks based on CSI. Similarly, \cite{10279804} proposed an RL-based beam selection method using a grid of beams (GoBs) technique, tested in the O-RAN framework as rApp to enhance energy efficiency.

Despite these advancements, current research on PMI selection does not fully address it as a multi-objective problem that balances SE with interference mitigation. Additionally, there is a lack of studies focusing on PMI selection within the O-RAN framework. While existing works underscore the importance of interference management in network optimization, they do not specifically tackle the challenge of PMI selection in a coordinated and interference-aware manner.

\section{System model}  
\subsection{Network model}

This work considers a network system of $K$ cells with UEs randomly distributed amongst the cells. This system employs an FDD-based 5G NR for downlink (DL) spatial multiplexing for data streaming to the UEs. The BS sends CS-RS, and based on the CS-RS, the UE sends a feedback response of the CSI to the BS. The CSI contains the CQI, RI, PMI, and other channel parameters (throughput, RSRP, etc). A frequency-selective PMI Feedback in FDD is considered where the UE sends different PMIs for different frequency subbands from the codebook.

\subsection{Codebook and PMI selection}
A BS and UE in a cell have $N_{t}$ and $N_{r}$ numbers of antennas respectively.  This multi-antenna system comprises $N_1$ and $N_2$ horizontal and vertical antennas with cross-polarization. The total number of transmit antennas becomes 2$N_1N_2$.

Multiple UEs \( u \in \mathcal{U}_k \) are served by a BS \( k \). The received signal \( \mathbf{y}_u(t) \in \mathbb{C}^{N_r \times 1} \) for the \( u \)-th UE can be expressed as:

\begin{equation}
\mathbf{y}_u(t) = \mathbf{H}_u(t) \mathbf{W}_u(t) \mathbf{s}_u(t) + \sum_{j \neq u} \mathbf{H}_u(t) \mathbf{W}_j(t) \mathbf{s}_j(t) + \mathbf{n}_u(t)
\label{eqn1}
\end{equation}

Where $\mathbf{H}_u(t) \in \mathbb{C}^{N_r \times N_t}$ is the channel matrix between the BS and the $u$-th UE at time t, $\mathbf{W}_u(t) \in \mathbb{C}^{N_t \times r_u}$ is the precoding matrix (PM) used for the \( u \)-th UE (selected based on PMI and RI feedback), where \( r_u \) is the rank (number of spatial streams) for the UE, $\mathbf{s}_u(t) \in \mathbb{C}^{r_u \times 1}$  is the transmitted signal vector for UE \( u \) with rank \( r_u \). The term \( \sum_{j \neq u} \mathbf{H}_u(t) \mathbf{W}_j(t) \mathbf{s}_j(t) \) represents interference from other UEs served by the same BS and \( \mathbf{n}_u(t) \in \mathbb{C}^{N_r \times 1} \) is the additive white Gaussian noise with zero mean and variance \( \sigma^2 \).

As illustrated in (\ref{eqn1}), $\mathbf{W}_u$ is the PM selected for the UE $(u)$ at time t.The beam steering technique adopted is based on selecting one PM from a list of matrices called codebook. This technique exploits the 3GPP codebook Type I defined in \cite{3gpp38213} which contains a set of PMs \( \mathbf{W}_{r,j} \), where \( r \) is the rank and \( j \) is the PMI index.

\subsection{Codebook Structure}\label{code}

The codebook \( \mathcal{W} \) contains PMs designed to optimize transmission for different ranks and channel conditions. For a UE with rank \( r \), the available PMs \( \mathbf{W}_{r,j} \) are designed to map \( r \) spatial streams onto the BS’s \( N_t \) antennas.

For rank \( r \), the PM \( \mathbf{W}_{r,j} \) comes from a predefined set of matrices based on the channel's spatial characteristics.

\begin{itemize}
    \item For \textbf{RI = 1} (rank 1), a set of PMs \( \mathbf{W}_{1,j} \), where \( j \in \{1, 2, \dots, J_1\} \).
    \item For \textbf{RI = 2} (rank 2), a set of PMs \( \mathbf{W}_{2,j} \), where \( j \in \{1, 2, \dots, J_2\} \).
\end{itemize}

\subsection{ PMI selection method (Follow PMI)} \label{sec}

In PMI selection, the BS decides whether the suggested PMI values from the UE are optimal for the network scenario or whether the BS must assign the UE a different PMI based on the BS configuration and optimization analysis. The BS makes this selection based on CSI feedback provided at each time slot, which includes PMIs for different frequency subbands, and chooses the PMI that maximizes the SE for each subband.


The SE \( C_{r,j,u}(t) \) for a UE \( u \) with rank \( r \) and PMI \( j \) at time \( t \) is given by:

\begin{equation}
C_{r,j,u}(t) = r \cdot \log_2 \left( 1 + \text{SNR}_{r,j,u}(t)\right)
\label{eqqn}
\end{equation}

and
$\text{SNR}_{r,j,u}(t) = \frac{\left\| \mathbf{H}_u(t) \mathbf{W}_{r,j} \right\|^2}{\sigma^2}$

\subsection{BS Selection of PMI}

At each time slot \( t \), the BS receives multiple potential PMIs from each UE and selects the PMI that maximizes the spectral efficiency (SE) $j_{u}^* = \arg \max_{r,j} C_{r,j,u}(t)$.


The BS then uses this PMI to select the corresponding PM \( \mathbf{W}_{r,j_{u}^*(t)} \) for the UE during that time slot.


The received signal for each UE \( u \) after PMI selection then becomes

\begin{equation}
\mathbf{y}_u(t) = \mathbf{H}_u(t) \mathbf{W}_{r_u,j_{u}^*} \mathbf{s}_u(t) + \sum_{j \neq u} \mathbf{H}_u(t) \mathbf{W}_{r',j'} \mathbf{s}_j(t) + \mathbf{n}_u(t)
\label{finalequation}
\end{equation}

Where \( r_u, j_{u}^* \) are the rank and PMI selected for UE \( u \) by the BS and \( r', j' \) are the rank and PMI for interfering UEs.


\subsection{Problem Definition}

The problem involves optimizing the selection of PMI values for a set of UEs within the network. As discussed in Section \ref{sec}, the BS is constrained to selecting a PMI that improves the SE of individual UEs, without considering network-wide optimization. Such an approach could lead to sub-optimal overall network performance. The goal is to choose PMI values that maximize SE while minimizing interference and ensuring efficient use of Physical Resource Blocks (PRBs), aided by the CSI feedback provided by the UEs. The objective function is defined as
\begin{align}
\max_{j_u^*} \sum_{u \in \mathcal{U}} \sum_{r,j} \left( \gamma_u - \lambda |\gamma_u - \gamma_{target}| \right)
\end{align}

\text{subject to:}

\[\sum_{j \neq u} \mathbf{H}_u(t) \mathbf{W}_{r',j'} \mathbf{s}_j(t) \leq I_{\text{max},u} + \partial, \quad \forall u \in \mathcal{U}\]

\[
\left| \sum_{u \in \mathcal{U}} PRB_{\text{used},u} - \wp \right| \leq \delta \]
where $\gamma_u$ is the SE of UE as in eq. \ref{eqqn} and $\gamma_{target}$ is the target SE. $\partial$ denotes the cost of interference experienced by the selected UEs. $\wp$ is the desired optimal PRB usage, used to penalize both underutilization and overutilization of PRBs with $\delta$ as the acceptable deviation from the optimal PRB usage.


The objective is to prioritize specific UEs, such as edge UEs or high-interference UEs, to improve their performance, especially in large-scale network scenarios with many UEs and cells.

This problem can be formulated as a Markov Decision Process (MDP), where the agent selects actions based on the network state to optimize the reward function, which accounts for both SE and interference minimization.
\subsection{Markov Decision Process (MDP)}

\begin{itemize}
    \item \textbf{State ($s_t$)}: The state at time $t$ captures key features of the cell’s performance. The state comprises of \text{mean CQI}($\nu$) of the non-prioritized UEs,  \text{maximum CQI}($cqi_{max}$) across all UEs, \text{cell interference} ($I_{k}$) experienced by UEs as defined in (\ref{eq7}), \text{mean PRB usage} ($\psi$) per UE, \text{average downlink throughput} ($avrg\_thrp$) in the cell, and \text{number of UEs} ($tot\_ues$) connected to the cell as shown in (\ref{eqnt}).
    
    \begin{equation}
    s_t = [\nu, cqi_{max}, I_{k, max}, \psi, avrg\_thrp, tot\_ues ] 
    \label{eqnt}
    \end{equation}
    These state parameters are normalized for stability and fast convergence. 
    \item \textbf{Action ($a_t$)}: The agent’s action at time $t$ consists of two decisions, selecting the prioritized group of UEs and assigning them PMI values. UEs are placed in three prioritization groups, 1). Edge UEs; those near the cell boundary. 2). High-interference UEs; those experiencing significant interference. 3). All UEs (apply the PMI selection uniformly).
 \begin{equation}   
 a_t = [J_1, J_2, \chi, \beta, \mathcal{U}_m, j_u^*]
 \label{act}
 \end{equation}
  where $J_1$ and $J_2$ are the lists of potential PMI values for RI 1 and RI 2 respectively as described in \ref{code}, $\chi$ denotes edge UEs (count of UEs with low RSRP that is below -100 dBm in the cell), $\beta$ for high interference UEs (sort UEs with the highest to lowest interference as described in \ref{eq7}), $\mathcal{U}_m$ is all UEs and $j_u^*$ is the actual PMI assigned to each UE.
    
    \item \textbf{Reward ($r_t$)}: The reward at time $t$ is designed to capture improvements in SE, interference reduction, and PRB usage efficiency. It is computed as:

     \begin{equation}
    r_t = (\gamma_u - \gamma_{target}) - \alpha \partial - |\Re - \wp|
    \label{rewaaad}
    \end{equation}
    where $\gamma_u$ is the SE of the UEs based on their CQI values and $\gamma_{target}$ is the target SE the network aims to achieve. $\alpha$ is a weighting factor for penalizing interference, $\partial$ denotes the cost associated with the interference experienced by the selected UEs (either edge UEs, high-interference UEs, or all UEs),  $\Re$ is the current total PRB usage in the downlink. 

\end{itemize}

\subsection{Advantage Actor-Critic (A2C)}
In the A2C method, we use two neural networks:
\begin{itemize}
    \item \textbf{Actor Network}: The actor is responsible for selecting the optimal action based on the current state. In this case, the action involves determining the PMI value and the UE prioritization strategy. The actor-network maps the state to a probability distribution over all possible actions.
    \item \textbf{Critic Network}: The critic evaluates the quality of the current state by estimating the value function $V(s_t)$, which represents the expected cumulative reward starting from state $s_t$. The critic provides feedback to the actor by helping the actor understand how good the action taken was in terms of expected future rewards.
\end{itemize}

In the simulation analysis, we compared our proposed method (Inter-A2C) with the Follow PMI selection method and A2C. The A2C was trained with the same state space and reward function. However, there was no UE prioritization and resource utilization. The action space for the A2C method is described in (\ref{action})
\begin{equation}   
 a_t = [J_1, J_2]
 \label{action}
 \end{equation}

\subsection{Cell selection for optimization}

For $\mathcal{U}_k$ UEs distributed in cell $k$, the total interference ($i_{u,k}$) a UE $u \in \mathcal{U}_k$ experiences in cell $k \in K$ comprises all the interference ($\iota_{\eta}$) received from the neighboring cells $N_{cell}^{k}$. The interference at the UE level and the entire cell is illustrated in (\ref{eq7}).
\begin{equation}
    i_{u, k} = \sum_{\eta = 1}^{N_{cell}^{k}} \iota_{\eta} \quad
    I_{k} = \sum_{u = 1}^{\mathcal{U}_k} \sum_{\eta = 1}^{N_{cell}^{k}}\iota_{\eta}.
    \label{eq7}
\end{equation}

where $U_k$ is the total number of UEs in BS $k$, and $\mathcal{U}_k \subseteq \mathcal{U}$, the total set of UEs in the network. This formulation enables the agent to determine which cells to optimize at each time step, implying that the agent will prioritize cells with the maximum total interference, i.e., those with $\arg \max_{k} I_{k}(t)$, where $I_k(t)$ is the total interference in cell $k$ at time step $t$. Only the cell with the highest interference is used at each time stamp.

\section{Performance and analysis}
\subsection{Simulation setup and training}

The proposed scheme is implemented in an O-RAN simulator developed by Telecom Italia (TIM) based on the Urban Macro (UMa) scenario defined in 3GPP TR 38.901. Details of this simulator are presented in \cite{10555357}. The simulation parameters for this scenario are described in Table \ref{tab} where the UEs are stationary. These models are deployed in docker containers and run on a GPU cluster. Each simulation takes 10ms on average, while feedback response from the UEs occurs every 1ms.

\begin{table}[htbp!]
\centering
\renewcommand{\arraystretch}{1.1}
\small
\caption{Simulation parameters}
\begin{tabular}{|p{0.55\columnwidth}|p{0.35\columnwidth}|}
\hline
\textbf{Parameters} & \textbf{Values} \\ \hline
Channel Bandwidth & 10 MHz \\ \hline
DL carrier frequency & 3.7 GHz \\ \hline 
Inter-Site Distance & 500m \\ \hline
DL Data Traffic & Full Buffer \\ \hline
Physical cell ID (PCI) & 57 \\ \hline
Number of UEs & 570  \\ \hline
Number of Transceivers  & 8  \\ \hline
Number of Antenna elements at the receiver &  2  \\ \hline
BS transmission power & 43dbm \\ \hline
Propagation Model & 3GPP TR 38.900 Urban Macro \\ \hline
Traffic Type & DL UDP (Full buffer)  \\ \hline
\end{tabular}
\label{tab}
\end{table}

\begin{figure}[htbp!]
    \centering
    \includegraphics[width=8cm]{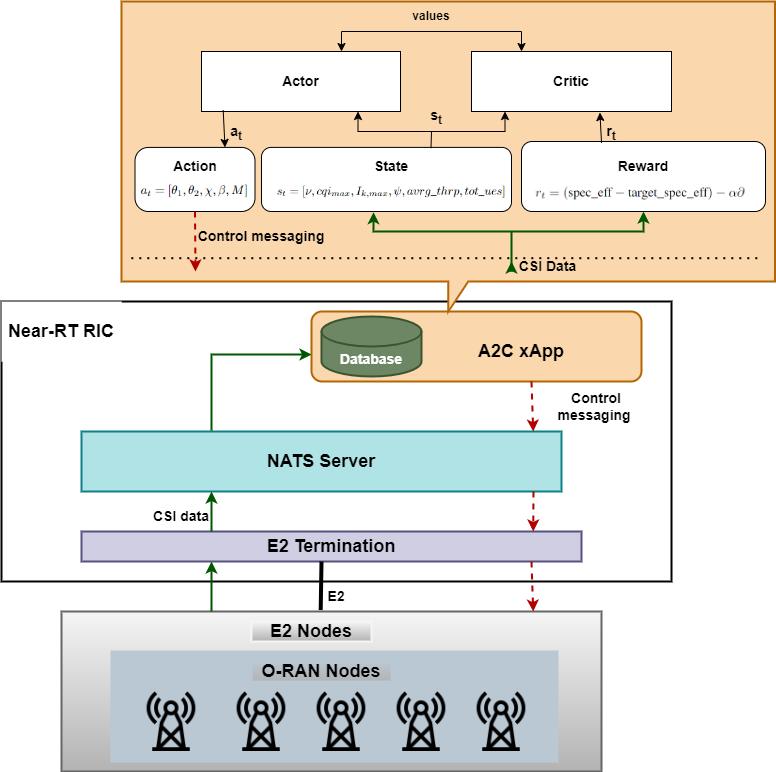}
    \caption{Simulation setup and training procedure}
    \label{workflow}
\end{figure}

A detailed simulation setup for our proposed scheme is illustrated in Fig. \ref{workflow}.  

\begin{figure}[htbp!]
    \centering
    \includegraphics[width=8cm]{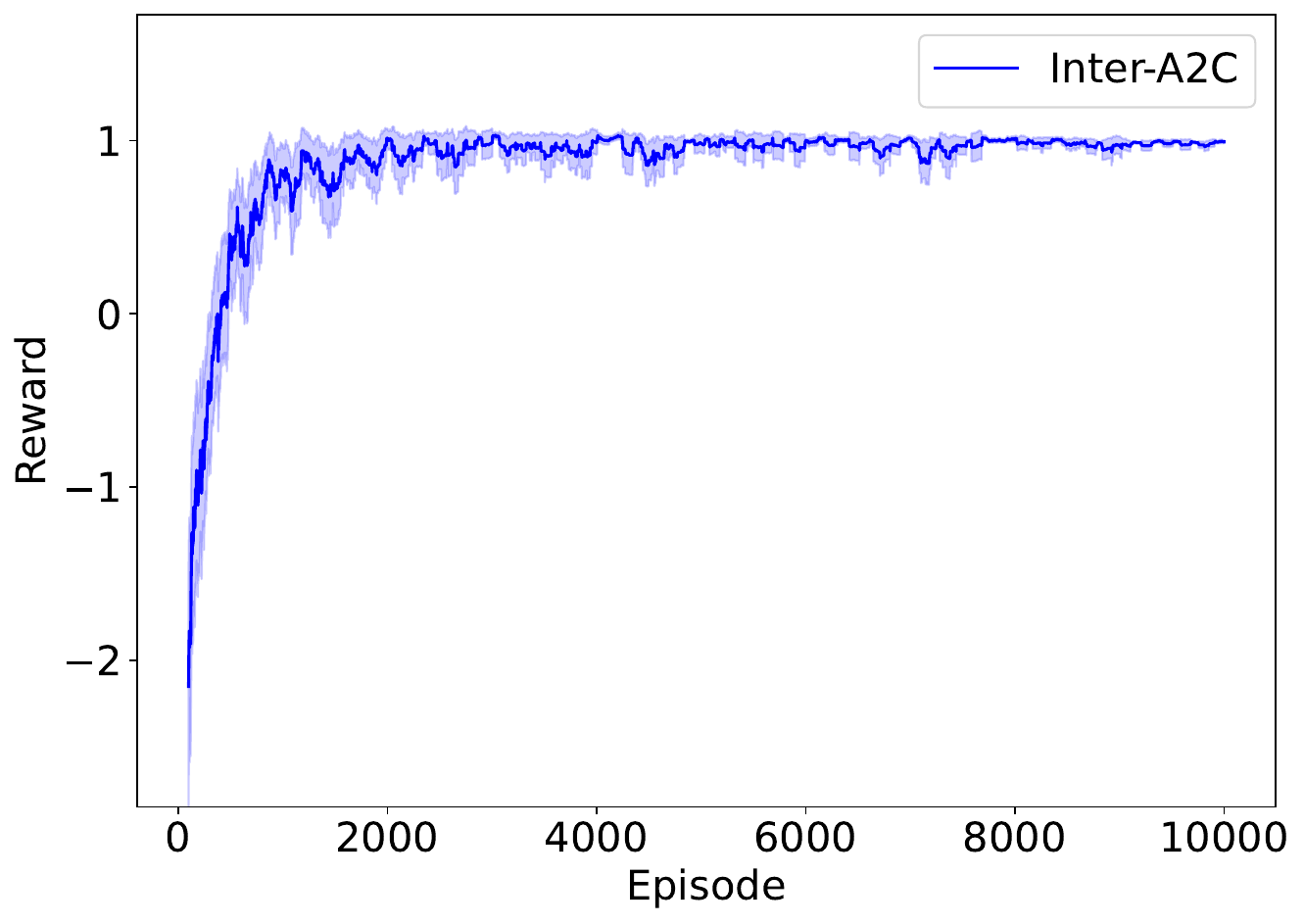}
    \caption{Rewards of the training of the A2C model}
    \label{rew}
\end{figure}

At every TTI, UEs send their CSI through the E2 interface. NATS serves as the routing manager, where a subscribed message is sent to the xApp (A2C) agent. The xApp is subscribed to receive CSI from UEs. We then compute the interference cells as described in eq. \ref{eq7}. In the cell optimization method, the agent selects the cell with the highest interference level. 
The A2C implementation was done using the Stable Baselines 3 defined in \cite{raffin2021stable}. As described in eq. \ref{rewaaad}, we defined the following parameters as follows:
\begin{itemize}
    \item \text{target\_spec\_eff}: 2.5
    \item $\alpha $: 0.7 (putting a little more emphasis on interference)
    \item $\wp$: 0.85 (85\%) of resource utilization
\end{itemize}
The CQI lookup table defined by the 3GPP TS 38.214, Table 5.2.2.1-2 and also described in \cite{basilashvili2017study} is used in obtaining the value of \text{target\_spec\_eff}. Hence the SE used in this work is computed as such.
  At every TTI, the UEs send the CSI information to the BSs. The agent receives these messages through the E2 interface. At each training episode, the agent selects the UE/UEs and the associated PMI values and sends a control message to the BS (PCI). The UE/UEs are then assigned the suggested PMI values and return the corresponding states and the reward is computed. The same or different cell could be selected in the next episode based on the interference levels. The convergence curve in Fig. \ref{rew} depicts the reward values obtained by A2C over 10000 episodes. 

\subsection{Results and analysis}

\begin{figure}[htbp!]
    \centering
    \includegraphics[width=8cm]{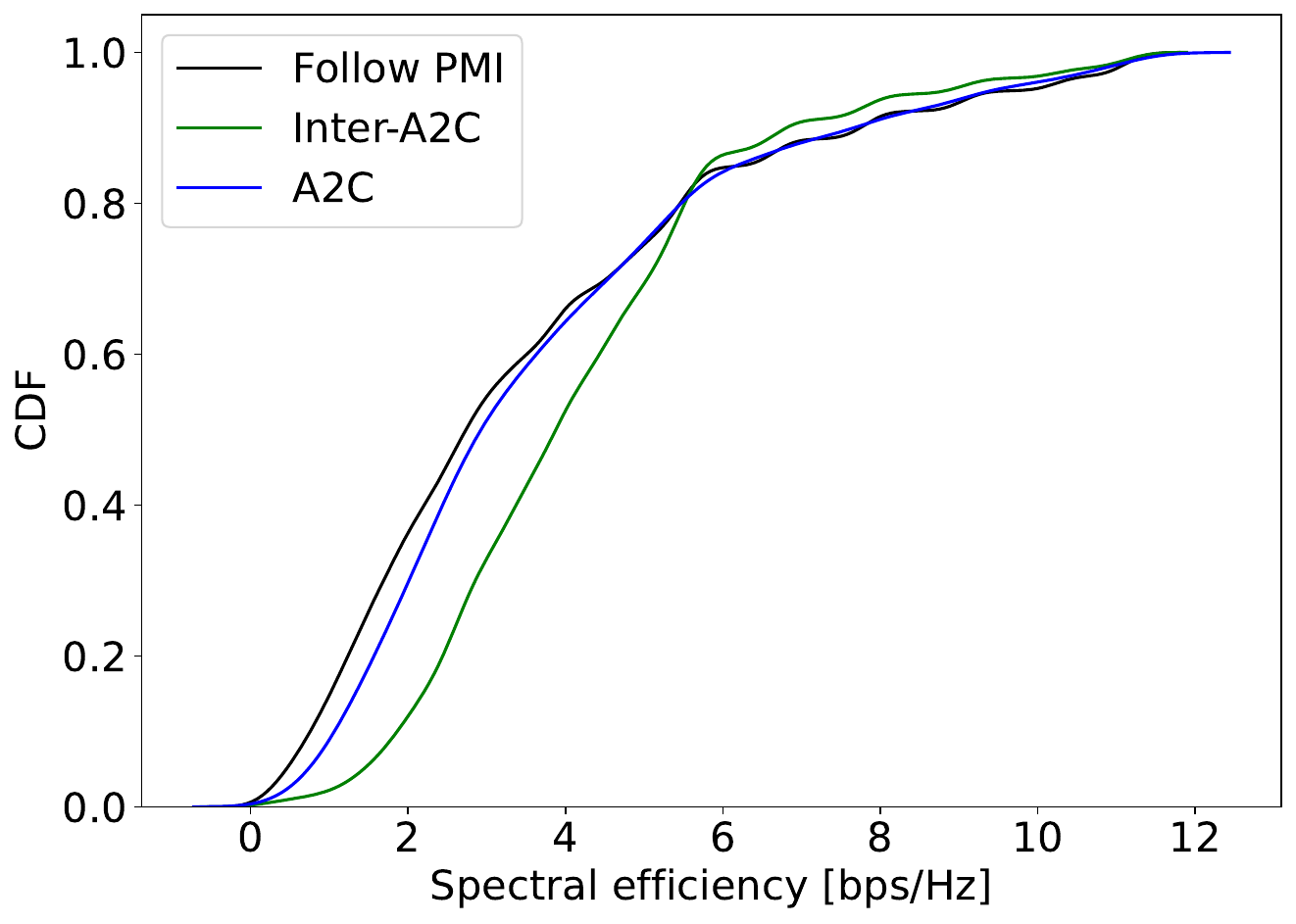}
    \caption{CDF of the spectral efficiency of the network}
    \label{cdf_spec}
\end{figure}

The CDF of the spectral efficiency (SE) for the three methods is presented in Fig. \ref{cdf_spec}. Our proposed method has significantly improved the SE of the network by refining the decision-making process through prioritization of PMI selection and addressing UEs with high interference and those at the cell edge. The Inter-A2C algorithm focused on optimizing PMI values and associated rewards, with prioritization playing a key role in improving SE. Although the A2C method showed improvements in SE, our proposed model further enhances performance by giving special consideration to UEs experiencing high interference and those located at the cell edge, leading to more refined and targeted decisions.

\begin{figure}[htbp!]
    \centering
    \includegraphics[width=8cm]{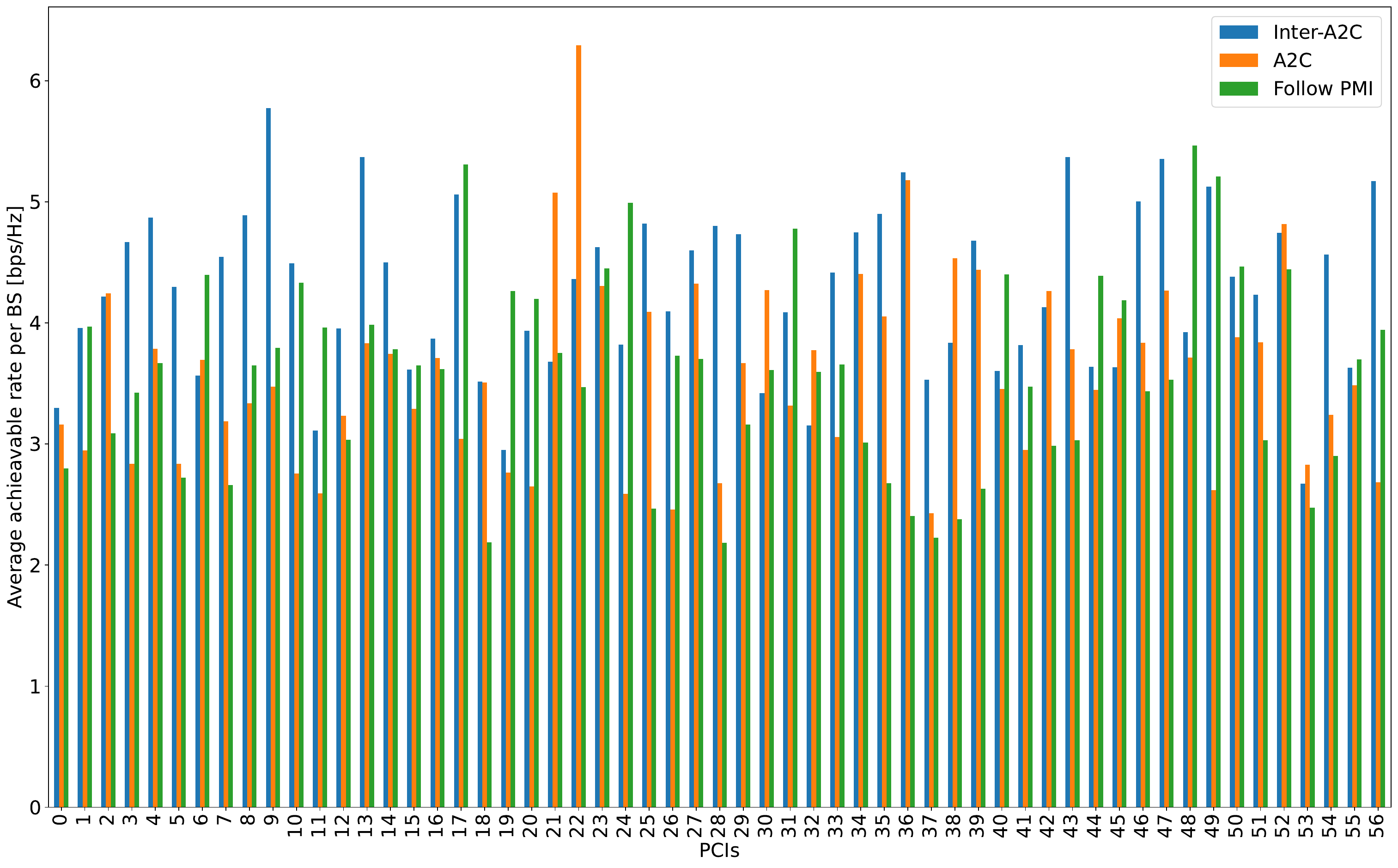}
    \caption{Average spectral efficiency per cell}
    \label{spec_dist}
\end{figure}

To better understand the behavior of the different methods in individual cells, we analyzed the average SE per cell, as shown in Fig. \ref{spec_dist}. While the A2C method provided the highest SE value recorded in a single cell, our proposed method (Inter-A2C) outperforms A2C regarding overall network performance. This indicates that although A2C excelled in specific scenarios, such as PMI assignments, our method ensures more consistent and balanced SE across the entire network.

\begin{figure}[htbp!]
    \centering
    \includegraphics[width=8cm]{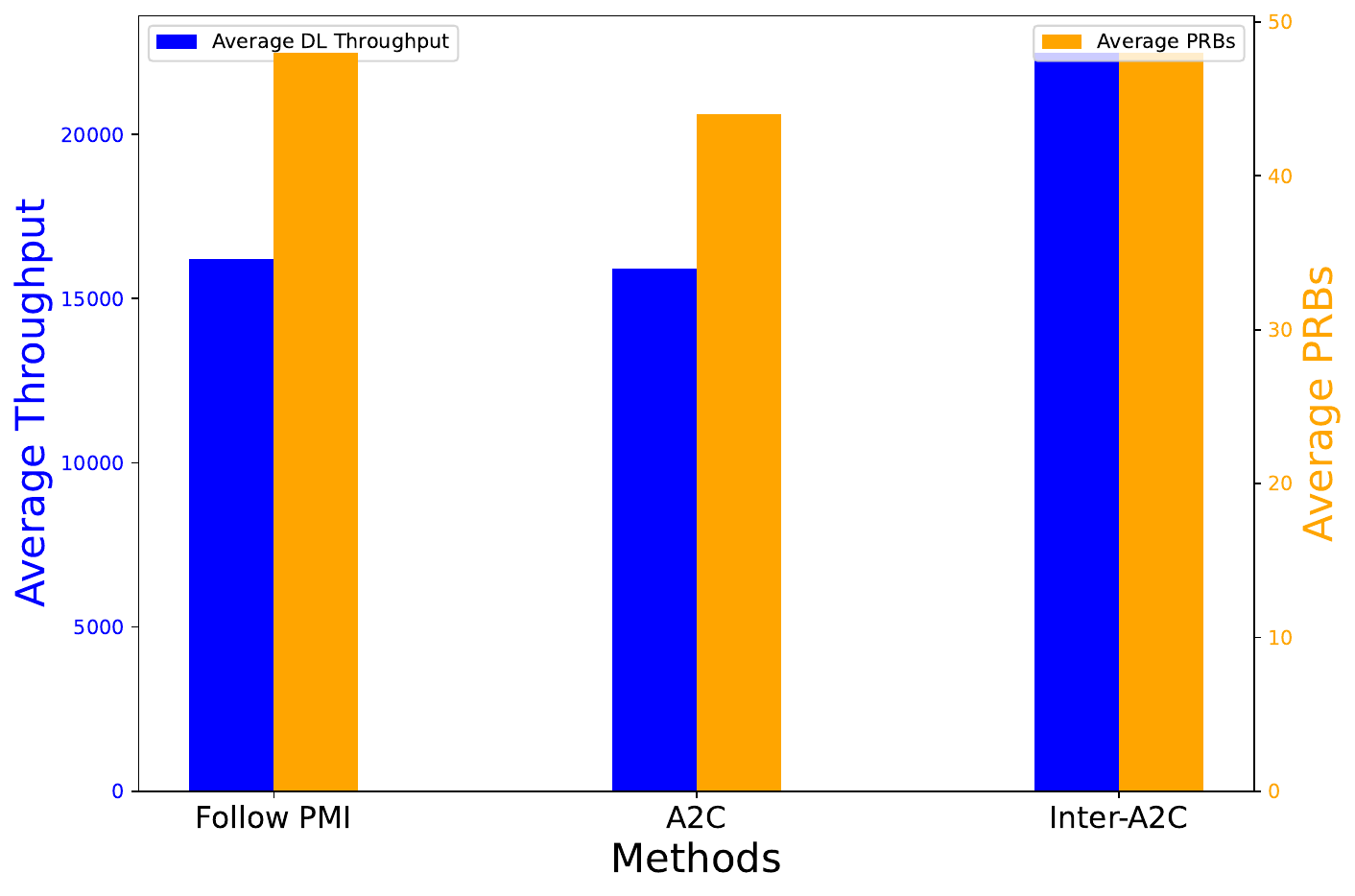}
    \caption{Average throughput and PRBs utilization in the network}
    \label{thrp}
\end{figure}

\begin{figure}[htbp!]
    \centering
    \includegraphics[width=8cm]{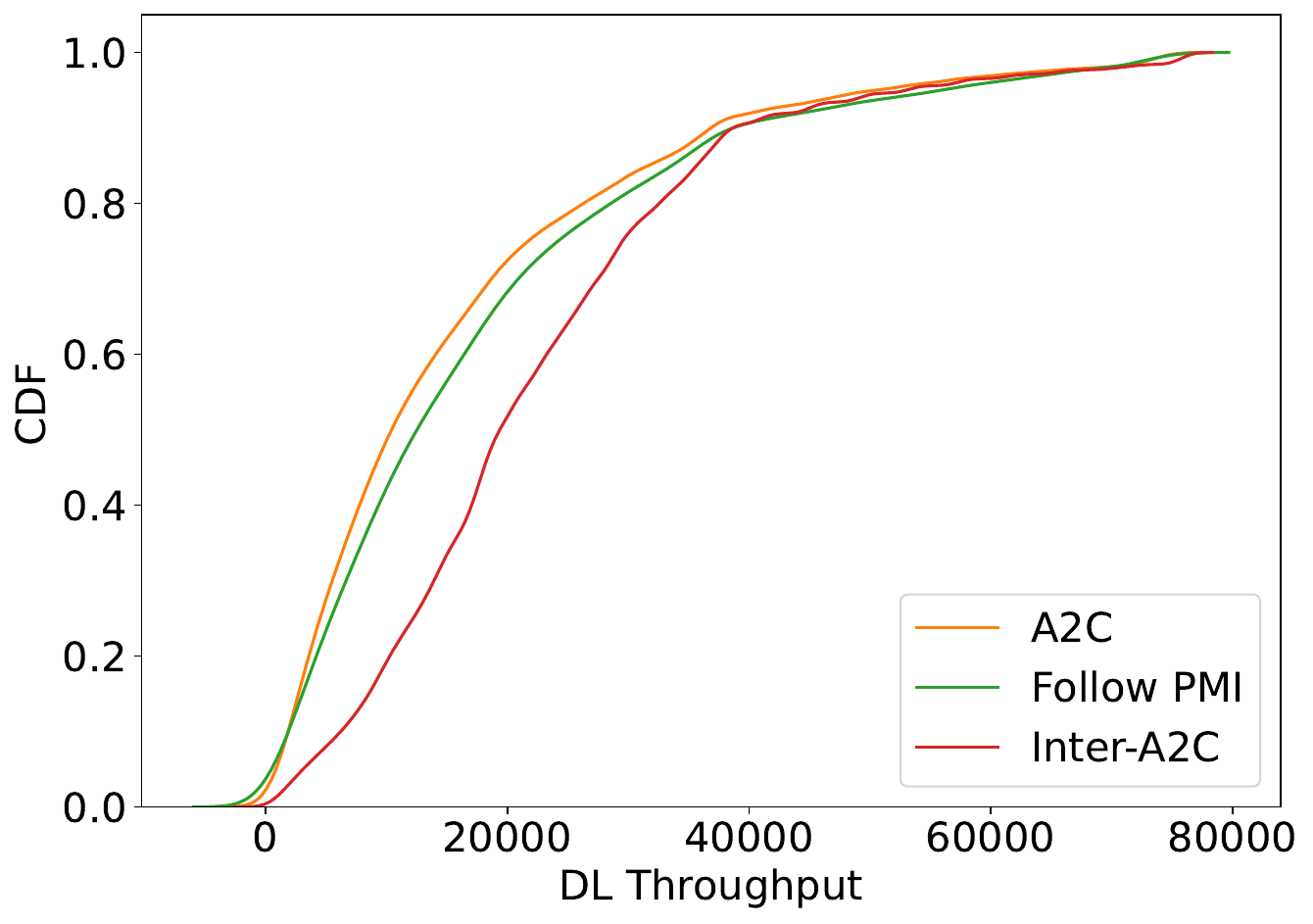}
    \caption{CDF of the throughput for all the methods}
    \label{cdfthrp}
\end{figure}

In Fig. \ref{thrp}, the average throughput of our network is analyzed alongside the average PRB utilization. This illustrates the network-wide performance of the three methods considered in this work. The Follow PMI method resulted in the lowest performance because the cell selects the best PMI based solely on the UE feedback without considering the overall cell performance, leading to suboptimal network performance. On average, the throughput values obtained by each method reflect the performance that UEs can expect in the network. This is also evident in Fig. \ref{cdfthrp}

The A2C and Follow PMI methods exhibit high PRB utilization but low average throughput and SE. This suggests that cell-edge UEs and UEs experiencing high interference are not adequately prioritized, resulting in inefficiencies. In contrast, our proposed method ensures a more balanced utilization of network resources, leading to higher throughput and improved overall performance.

\begin{figure}[htbp!]
    \centering
    \includegraphics[width=8cm]{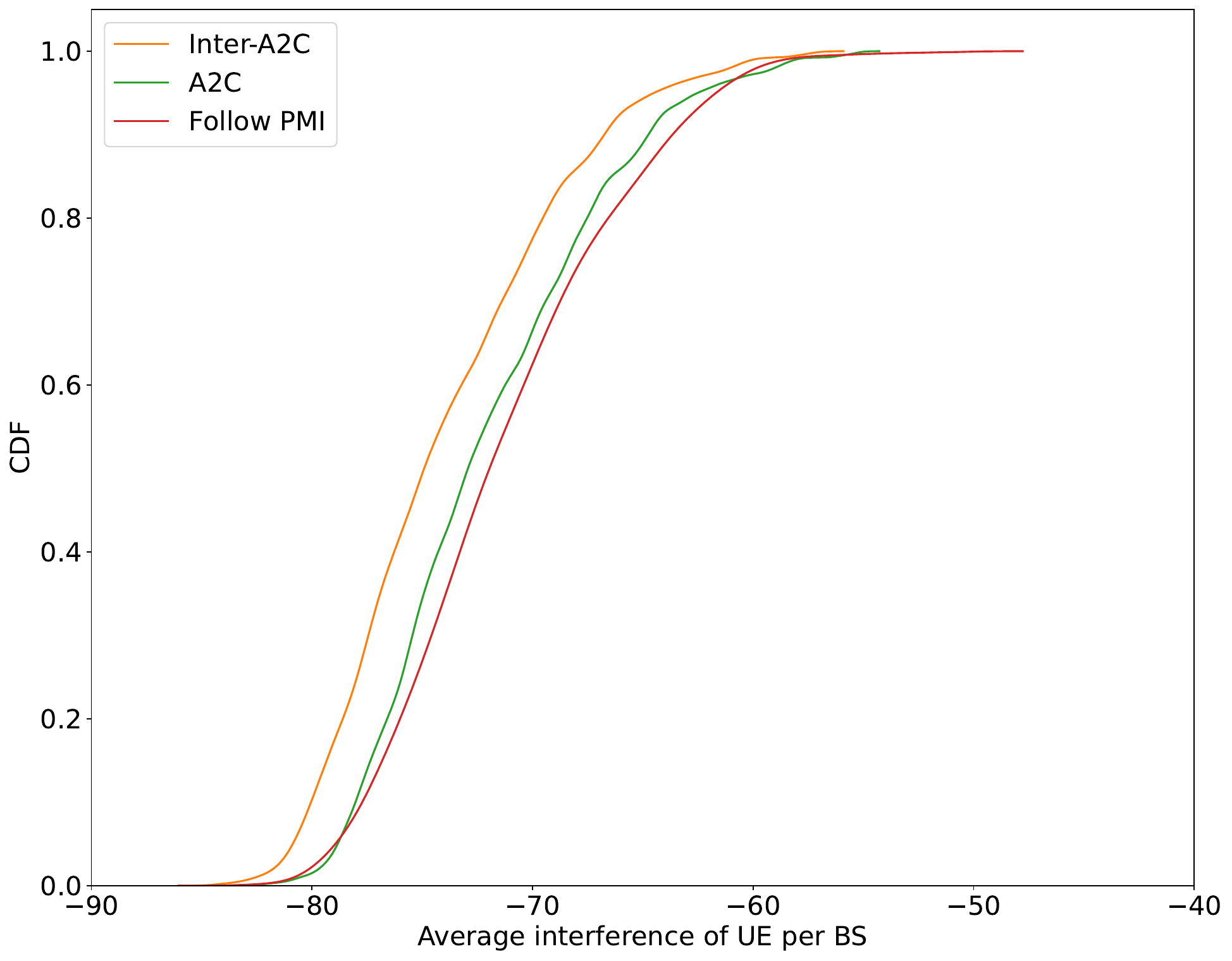}
    \caption{Avereage Interference }
    \label{aver_inter}
\end{figure}

Fig. \ref{aver_inter} provides an overview of the interference observed across different scenarios. Our proposed method results in a significant reduction in interference compared to the Follow PMI and A2C methods. In our environmental setup, each UE receives signals from 9 neighboring cells, leading to higher interference levels due to the 500m inter-site distance (ISD). However, as the graph shows, our method effectively mitigates interference by optimizing PMI selection and prioritizing UEs with high interference, resulting in improved network performance.

\section{Conclusion}
This work presents an innovative approach to PMI selection in 5G networks using an RL-based xApp designed for the O-RAN framework. The proposed method effectively balances spectral efficiency and interference reduction by leveraging an Interference-Aware Advantage Actor-Critic (Inter-A2C) model, addressing a critical gap in existing research. The experimental results demonstrate significant improvements in network performance, showcasing the potential of intelligent PMI management in real-world scenarios. This study underscores the importance of integrating AI/ML techniques within the O-RAN architecture for optimizing complex tasks such as PMI selection, paving the way for more adaptive and efficient network operations. 

Future work will explore further enhancements, including the integration of additional network parameters and real-time adaptation to dynamic network conditions. An interference prediction using AI/ML models for PMI selection is also an interesting direction to consider. Additionally, we plan to investigate cell clustering for multi-agent RL, where each agent optimizes a cluster of cells, rather than deploying a separate agent for each cell.

\section*{Acknowledgement}
  This work was jointly supported by Telecom Italia S.p.A. under the framework of the UniversiTIM program and the Italian National Inter-University Consortium for Telecommunications (CNIT). The research was partially funded by the European Union's Horizon Europe research and innovation program under the HORSE – Holistic, Omnipresent, Resilient Services for Future 6G Wireless and Computing Ecosystems – G.A. 101096342.

\ifCLASSOPTIONcaptionsoff
  \newpage
\fi



%
\bibliographystyle{IEEEtran}
\bibliography{IEEEabrv,references}

\begin{thebibliography}{10}
\providecommand{\url}[1]{#1}
\csname url@samestyle\endcsname
\providecommand{\newblock}{\relax}
\providecommand{\bibinfo}[2]{#2}
\providecommand{\BIBentrySTDinterwordspacing}{\spaceskip=0pt\relax}
\providecommand{\BIBentryALTinterwordstretchfactor}{4}
\providecommand{\BIBentryALTinterwordspacing}{\spaceskip=\fontdimen2\font plus
\BIBentryALTinterwordstretchfactor\fontdimen3\font minus \fontdimen4\font\relax}
\providecommand{\BIBforeignlanguage}[2]{{%
\expandafter\ifx\csname l@#1\endcsname\relax
\typeout{** WARNING: IEEEtran.bst: No hyphenation pattern has been}%
\typeout{** loaded for the language `#1'. Using the pattern for}%
\typeout{** the default language instead.}%
\else
\language=\csname l@#1\endcsname
\fi
#2}}
\providecommand{\BIBdecl}{\relax}
\BIBdecl

\bibitem{zhang2019towards}
S.~Zhang, Y.~Wang, and W.~Zhou, ``Towards secure 5g networks: A survey,'' \emph{Computer Networks}, vol. 162, p. 106871, 2019.

\bibitem{rao20215g}
L.~Rao, M.~Pant, L.~Malviya, A.~Parmar, and S.~V. Charhate, ``5g beamforming techniques for the coverage of intended directions in modern wireless communication: in-depth review,'' \emph{International Journal of Microwave and Wireless Technologies}, vol.~13, no.~10, pp. 1039--1062, 2021.

\bibitem{3gpp38213}
3GPP, ``5g;nr;physical layer procedures for control (release 15),'' \emph{3GPP TS 38.213 version 15.8.0}, 2020.

\bibitem{8938771}
F.~B. Mismar, B.~L. Evans, and A.~Alkhateeb, ``Deep reinforcement learning for 5g networks: Joint beamforming, power control, and interference coordination,'' \emph{IEEE Transactions on Communications}, vol.~68, no.~3, pp. 1581--1592, 2020.

\bibitem{10224533}
J.-S. Sheu, C.-K. Huang, and C.-L. Tsai, ``Joint beamforming, power control, and interference coordination: A reinforcement learning approach replacing rewards with examples,'' \emph{IEEE Access}, vol.~11, pp. 88\,854--88\,868, 2023.

\bibitem{unlercsen2016direction}
M.~F. {\"U}NLER{\c{s}}EN and E.~Yaldiz, ``Direction of arrival estimation by using artificial neural networks,'' in \emph{2016 European Modelling Symposium (EMS)}.\hskip 1em plus 0.5em minus 0.4em\relax IEEE, 2016, pp. 242--245.

\bibitem{9259040}
P.~Ramezanpour and M.-R. Mosavi, ``Two-stage beamforming for rejecting interferences using deep neural networks,'' \emph{IEEE Systems Journal}, vol.~15, no.~3, pp. 4439--4447, 2021.

\bibitem{kao2018ai}
W.-C. Kao, S.-Q. Zhan, and T.-S. Lee, ``Ai-aided 3-d beamforming for millimeter wave communications,'' in \emph{2018 International Symposium on Intelligent Signal Processing and Communication Systems (ISPACS)}.\hskip 1em plus 0.5em minus 0.4em\relax IEEE, 2018, pp. 278--283.

\bibitem{8114345}
A.~Ali, N.~González-Prelcic, and R.~W. Heath, ``Millimeter wave beam-selection using out-of-band spatial information,'' \emph{IEEE Transactions on Wireless Communications}, vol.~17, no.~2, pp. 1038--1052, 2018.

\bibitem{9034044}
M.~S. Sim, Y.-G. Lim, S.~H. Park, L.~Dai, and C.-B. Chae, ``Deep learning-based mmwave beam selection for 5g nr/6g with sub-6 ghz channel information: Algorithms and prototype validation,'' \emph{IEEE Access}, vol.~8, pp. 51\,634--51\,646, 2020.

\bibitem{9681382}
M.~Zecchin, M.~B. Mashhadi, M.~Jankowski, D.~Gündüz, M.~Kountouris, and D.~Gesbert, ``Lidar and position-aided mmwave beam selection with non-local cnns and curriculum training,'' \emph{IEEE Transactions on Vehicular Technology}, vol.~71, no.~3, pp. 2979--2990, 2022.

\bibitem{love2003grassmannian}
D.~J. Love, R.~W. Heath, and T.~Strohmer, ``Grassmannian beamforming for multiple-input multiple-output wireless systems,'' \emph{IEEE transactions on information theory}, vol.~49, no.~10, pp. 2735--2747, 2003.

\bibitem{5379063}
J.~Wang, Z.~Lan, C.-S. Sum, C.-W. Pyo, J.~Gao, T.~Baykas, A.~Rahman, R.~Funada, F.~Kojima, I.~Lakkis, H.~Harada, and S.~Kato, ``Beamforming codebook design and performance evaluation for 60ghz wideband wpans,'' in \emph{2009 IEEE 70th Vehicular Technology Conference Fall}, 2009, pp. 1--6.

\bibitem{zhang2018joint}
K.~Zhang, W.~Tan, G.~Xu, C.~Yin, W.~Liu, and C.~Li, ``Joint rrh activation and robust coordinated beamforming for massive mimo heterogeneous cloud radio access networks,'' \emph{IEEE Access}, vol.~6, pp. 40\,506--40\,518, 2018.

\bibitem{bjornson2017massive}
E.~Bj{\"o}rnson, J.~Hoydis, L.~Sanguinetti \emph{et~al.}, ``Massive mimo networks: Spectral, energy, and hardware efficiency,'' \emph{Foundations and Trends{\textregistered} in Signal Processing}, vol.~11, no. 3-4, pp. 154--655, 2017.

\bibitem{9605010}
J.~Akhtar, K.~Saija, N.~Ravi, S.~Nethi, and S.~Chaudhuri, ``Machine learning-based prediction of pmi report for dl-precoding in 5g-nr system,'' in \emph{2021 IEEE 4th 5G World Forum (5GWF)}, 2021, pp. 105--110.

\bibitem{9762865}
V.~Ranjbar, A.~Girycki, M.~A. Rahman, S.~Pollin, M.~Moonen, and E.~Vinogradov, ``Cell-free mmimo support in the o-ran architecture: A phy layer perspective for 5g and beyond networks,'' \emph{IEEE Communications Standards Magazine}, vol.~6, no.~1, pp. 28--34, 2022.

\bibitem{dreifuerst2023ml}
R.~M. Dreifuerst and R.~W. Heath~Jr, ``Ml codebook design for initial access and csi type-ii feedback in sub-6ghz 5g nr,'' \emph{arXiv preprint arXiv:2303.02850}, 2023.

\bibitem{10279804}
Y.~Dantas, P.~E. Iturria-Rivera, H.~Zhou, M.~Bavand, M.~Elsayed, R.~Gaigalas, and M.~Erol-Kantarci, ``Beam selection for energy-efficient mmwave network using advantage actor critic learning,'' in \emph{ICC 2023 - IEEE International Conference on Communications}, 2023, pp. 5285--5290.

\bibitem{10555357}
R.~Ntassah, G.~Michele~Dell’Area, and F.~Granelli, ``User classification and traffic steering in o-ran,'' \emph{IEEE Open Journal of the Communications Society}, vol.~5, pp. 3581--3594, 2024.

\bibitem{raffin2021stable}
A.~Raffin, A.~Hill, A.~Gleave, A.~Kanervisto, M.~Ernestus, and N.~Dormann, ``Stable-baselines3: Reliable reinforcement learning implementations,'' \emph{Journal of Machine Learning Research}, vol.~22, no. 268, pp. 1--8, 2021.

\bibitem{basilashvili2017study}
G.~Basilashvili, ``Study of spectral efficiency for lte network,'' \emph{American Scientific Research Journal for Engineering, Technology, and Sciences (ASRJETS)}, vol.~29, no.~1, pp. 21--32, 2017.

\end{thebibliography}

%





\end{document}